\documentclass[aps,prb,twocolumn,superscriptaddress]{revtex4}
\usepackage{amsfonts}
\usepackage{amssymb}
\usepackage{amsmath}
\usepackage{graphicx}
\usepackage{epsfig}
\usepackage{color}
\usepackage{amsmath,bm}

\begin{document}

\title{Steady off-diagonal long-range order state in a half-filled %
dimerized Hubbard chain induced by a resonant pulsed field}
\author{X. Z. Zhang}
\affiliation{College of Physics and Materials Science, Tianjin Normal University, Tianjin
300387, China}
\author{Z. Song}
\email{songtc@nankai.edu.cn}
\affiliation{School of Physics, Nankai University, Tianjin 300071, China}

\begin{abstract}
We show that a resonant pulsed field can induce a steady superconducting
state even in the deep Mott insulating phase of the dimerized Hubbard model.
The superconductivity found here in the non-equilibrium steady state is due
to the $\eta $-pairing mechanism, characterized by the existence of the
off-diagonal long-range order (ODLRO), and is absent in the ground-state
phase diagram. The key of the scheme lies in the generation of the
field-induced charge density wave (CDW) state that is from the valence bond
solid. The dynamics of this state resides in the highly-excited subspace of %
dimerized Hubbard model and is dominated by a $\eta $-spin
ferromagnetic model. The decay of such long-living excitation is suppressed
because of energy conservation. We also develop a dynamical method to detect
the ODLRO of the non-equilibrium steady state. Our finding demonstrates that
the non-equilibrium many-body dynamics induced by the interplay between the
resonant external field and electron-electron interaction is an alternative
pathway to access a new exotic quantum state, and also provides an
alternative mechanism for enhancing superconductivity.
\end{abstract}

\maketitle

\section{Introduction}

Driving is not only a transformative tool to investigate complex many-body
system but also makes it possible to create non-equilibrium phase of quantum
matter with desirable properties \cite%
{Verstraete2009,Ichikawa2011,Eisert2015,Basov2017,Mor2017,Cavalleri2018,Ishihara2019}%
. It can significantly alter the microscopic behavior of strongly correlated
system and manifest a variety of collective and cooperative phenomena at the
macroscopic level. Spurred on by experiments in ultra-cold atomic gases, the
non-equilibrium strongly correlated systems have been the subject of intense
study over the last decade \cite%
{Greiner2002,Freericks2006,Bloch2008,Freericks2008,Aron2012,Aoki2014,Essler2016,Vidmar2016,Vasseur2016,Wang2016a,Zhang2020c,Rubio-Abadal2020,Moudgalya2020,Wei2021,Zhang2022}%
. Additionally, pump-probe spectroscopy offers a new avenue for the
exploration of available electronic states in correlated materials \cite%
{Perfetti2006}. Among them, the most striking is the discovery of
photoinduced transient superconducting behaviors in some high-$T_{c}$
cuprates \cite{Fausti2011,Hu2014,Kaiser2014} and alkali-doped fullerenes
\cite{Mitrano2016,Cantaluppi2018}. All these advances have revived interest
in the fundamental behavior of quantum systems away from equilibrium.
\begin{figure}[tbp]
\centering
\includegraphics[width=0.45\textwidth]{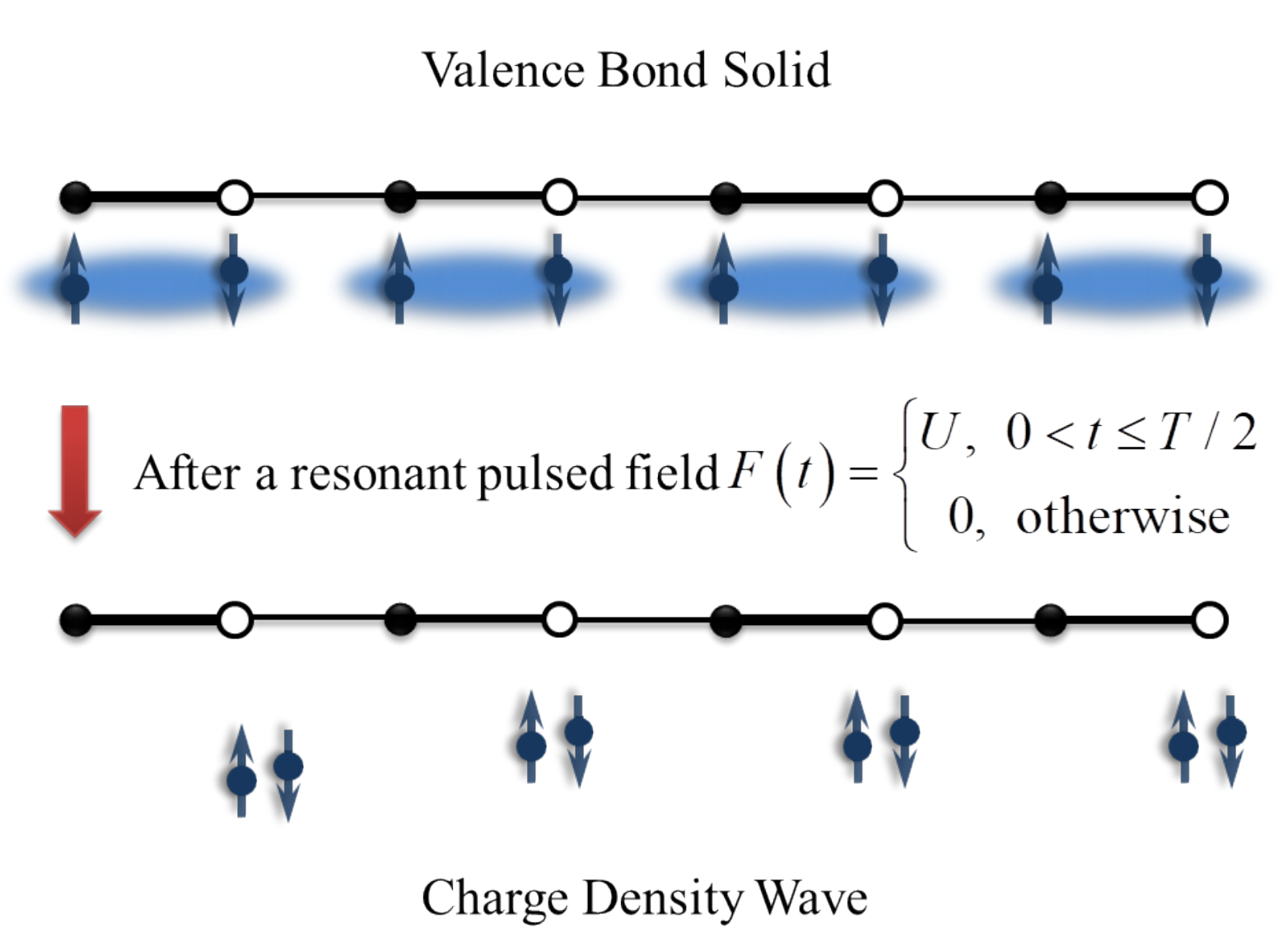}
\caption{Schematic illustration of the dynamical pairing process considered
in this work. The system is initialized in a dimerized Hubbard model at half
filling. The strong dimerization divides the whole $2N$ lattice into $N$
unitcells. In each unitcell, two electron spins form a spin $0$ singlet due
to the antiferromagnetic interaction, while not being entangled with the
spins of other unitcell. Hence, the gound state is a valence bond solid. The
resonant pulsed field $F$ plays the role in each unitcell individually such
that the CDW state can be generated after a period $T/2$ with $T=\protect\pi %
/\Delta $. Then the CDW state will evolve to an ODLRO state via doublon
diffusion, which is the key to realizing the non-equilibrium superconducting
phase.}
\label{fig_ill}
\end{figure}

Non-equilibrium control of quantum matter is an intriguing prospect with
potentially important technological applications \cite%
{Yonemitsu2008,Giannetti2016,Oka2019,Torre2021}. Experiments with various
materials and excitation conditions have witnessed phenomena with no
equilibrium analog or accessibility of chemical substitution, including
superconducting-like phases \cite%
{Fausti2011,Mitrano2016,Cavalleri2018,Suzuki2019}, charge density waves
(CDW) \cite{Stojchevska2014,Matsuzaki2014,Kogar2020} and excitonic
condensation \cite{Murotani2019}. Among various non-equilibrium protocols,
the generation of the $\eta $-pairing-like state possessing the off-diagonal
long-range order (ODLRO), originally proposed by Yang for the Hubbard model
\cite{Yang1989}, plays a pivotal role in which the existence of doublon and
holes facilitate the superconductivity \cite%
{Kitamura2016,Kaneko2019,Tindall2019,Fujiuchi2019,Peronaci2020,Fujiuchi2020,Ejima2020,Li2020,Kaneko2020,Zhang2021}%
. Therefore, how to stabilize a system in a non-equilibrium superconducting
phase with a long lifetime is a great challenge and is at the forefront of
current research. Besides, constructing a clear and simple physical picture
to realize the non-equilibrium superconducting phase for the experiment is
also the goal of on-going theoretical investigation.
\begin{figure*}[tbp]
\centering
\includegraphics[width=0.7\textwidth]{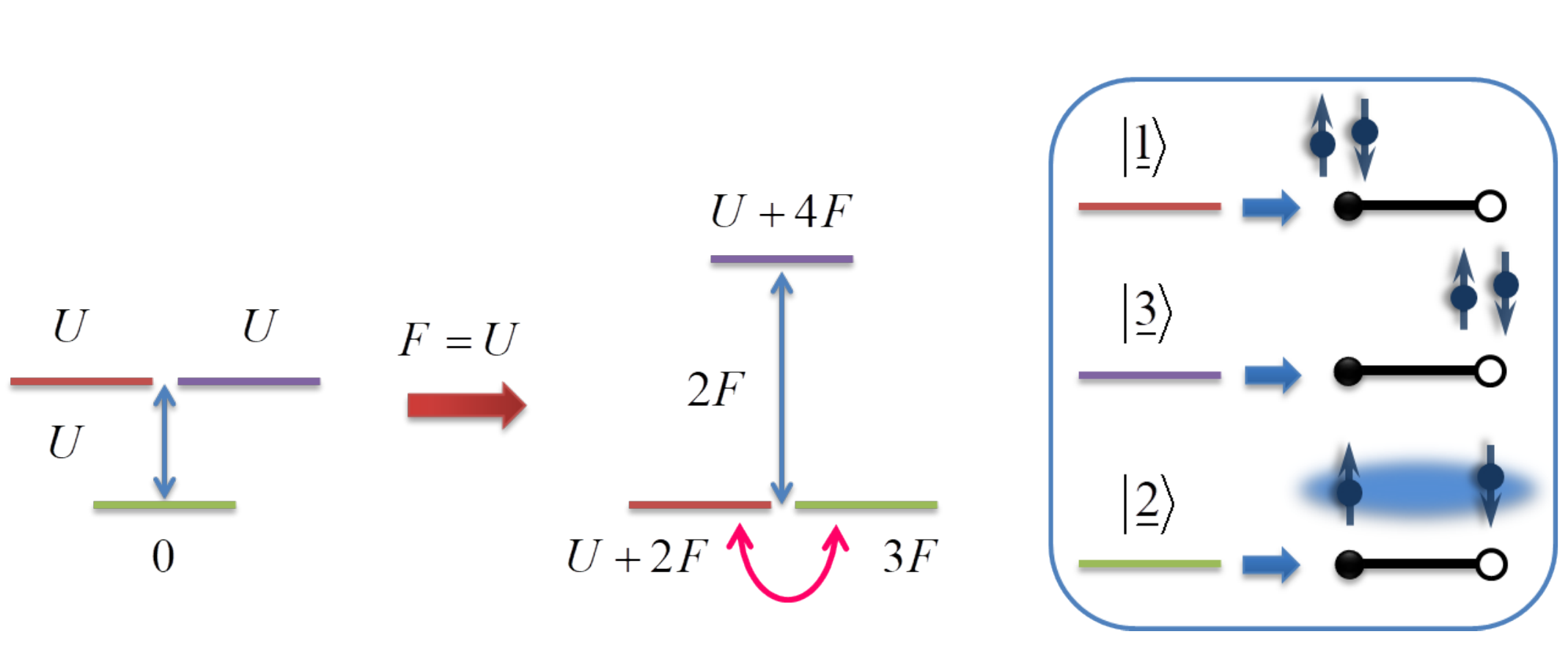}
\caption{Sketch of the resonant pairing
mechanism
in
the
two-site Hubbard
model at half-filling. The system can
be divided
into
two
subspaces labeled by the spin quantum numbers $s=0$,
and $s=1$.
We
focus on
the subspace with $s=0$. The double-occupied bases
are denoted
by
red and
purple lines and the green line indicates the
valence bond
state
that is the
GS when $U/\kappa\gg 1$. In the absence of
$F$, there
exists an
energy
difference of $U$ between such two types of
states. The
resonant $F$
places
the valence bond state and red
double-occupied state
on the same
energy
shell such that the kinetic term
allows a transfer
between these two
states.
The gap $2F$ prohibits the
tunneling from the
lower two states to
the upper
purple state and hence
protects the
formation of the CDW state in
the whole
lattice.}
\label{fig_illustration_resonant}
\end{figure*}

It is the aim of this paper to unveil the underlying mechanism of
superconductivity in a non-equilibrium matter. The core is how to excite a
Mott insulator to a pairing state (CDW state) within the highly excited
subspace. Then it evolves to a steady ODLRO state. To this end, we consider
a repulsive dimerized Hubbard model, in which the dimerization can control
the type of the ground state but does not change the magnetic correlation.
The strong dimerization can make the main component of the
anti-ferromagnetic ground state change from a Neel state to a valence bond
solid where the electrons belonging to the different unitcells are not
entangled with each other. This allows that the resonant pulsed field can
drive the spin singlet state to a double-occupied state in each unitcell so
that the CDW state is constructed in the entire lattice. The doublons and
holes can significantly enhance the conductivity of the system. Fig. \ref%
{fig_ill} illustrates this core dynamics of the proposed non-equilibrium
scheme. Note that non-resonant external field will also increase the
conductivity of the system, but will not form a CDW state with maximized
doublons and holes. This does not favor the superconductivity in the
subsequent dynamics. Due to the energy conservation, the system can stay in
the highly-excited subspace, which shares the same energy shell with the CDW
state, for a long time. The corresponding doublon dynamics can be fully
captured by the effective $\eta $-spin ferromagnetic model that can be
obtained through the virtual exchange of the particles. In this context,
such an effective Hamiltonian can drive the CDW state to a steady state
which distributes evenly in the lattice and possess the long-range $\eta $%
-spin correlation. This is the characteristic of the system entering the
non-equilibrium superconducting phase. By introducing the magnetic flux, we
further develop a method of detecting this kind of non-equilibrium phase of
matter based on the performance of the Loschmidt echo (LE). Specifically, the characteristic that LE shows
periodical behavior rather than a constant value around $1$ can be used to
detect whether the system is in the superconducting phase. It is hoped that
these results can motivate further studies of both the fundamental aspects
and potential applications of the non-equilibrium interacting system.

The remainder of this paper is organized as follows. In Sec. \ref{mechanism}%
, we first present the pairing dynamics induced by the resonant pulsed
field. Second, we explore the long-time dynamics of a single doublon and
extend the results to the multi-doublon case, which paves the way to achieve
the effective $\eta $-spin model and hence facilitates the understanding of
the steady ODLRO state. In Sec. \ref{scheme}, we propose a dynamical scheme
to excite the system into the non-equilibrium superconducting phase based on
the repulsive dimerized Hubbard model. Correspondingly, a
dynamical detection method is constructed to examine such a phase. Finally,
we conclude our results in Sec. \ref{summary}.
Some details of
our calculation are placed in the Appendix.

\section{Two simple models to elucidate the underlying mechanism}

\label{mechanism} Recently, much attention has been paid to the realization
of the superconductivity in the deep Mott insulator phase via
out-of-equilibrium dynamics, e.g., quench dynamics. The underlying mechanism
can be attributed to the $\eta $-pairing state induced by the external
field. From a deep level, however, such a statement is neither complete nor
the corresponding dynamic process is clear. In this section, we provide two
examples to unravel the field-induced superconductivity. Such two models
correspond to the two key parts of the entire dynamic process, namely the
pairing induced by the external field and the formation of the long-range
correlation via diffusion of doublon.

\subsection{pairing induced by a resonant tilted fled}

\begin{figure*}[tbp]
\centering
\includegraphics[width=0.9\textwidth]{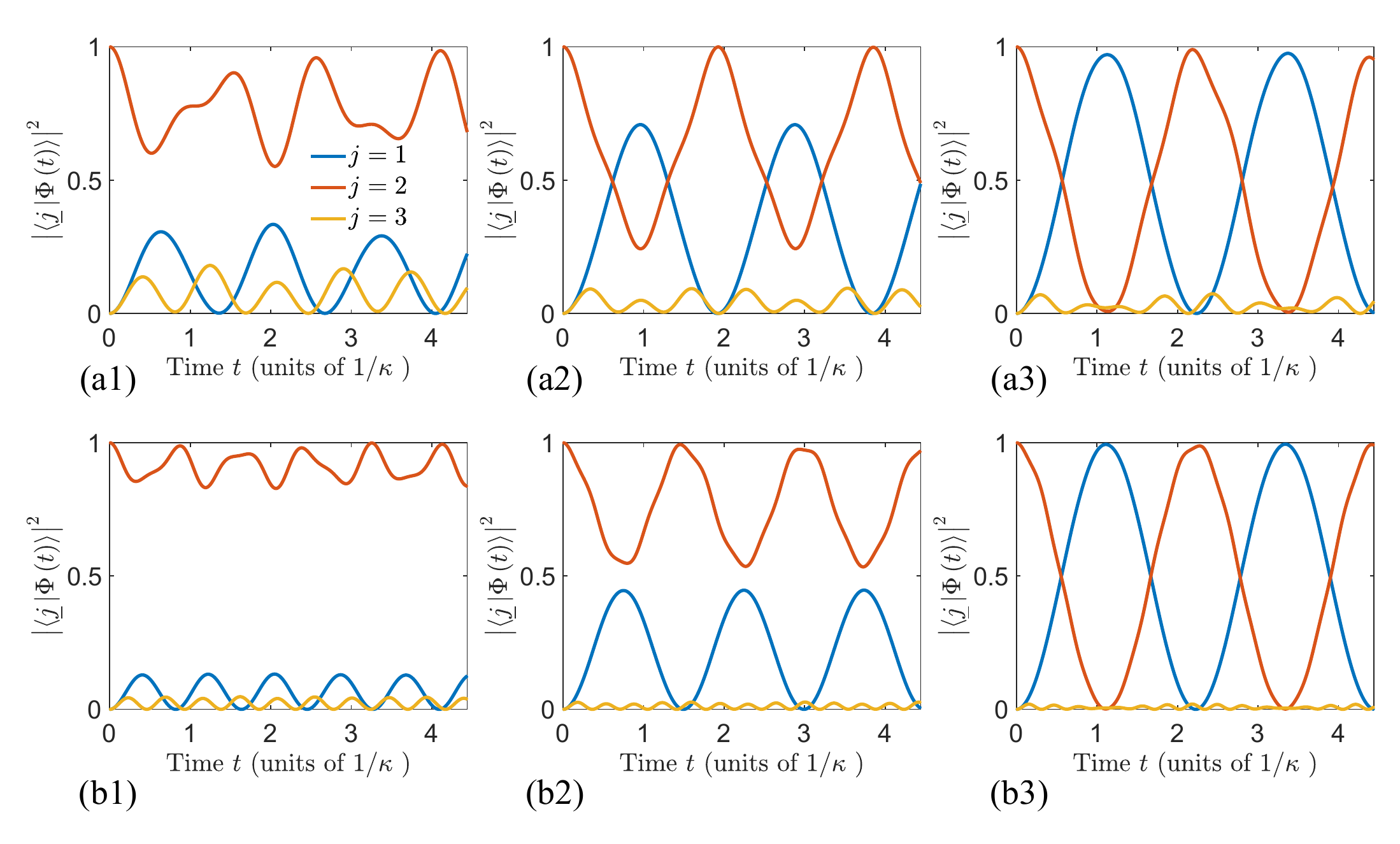}
\caption{Dynamical pairing of the $2$-site Hubbard model at half filling for
the different pulsed fields: (a1)-(b1) $F/U=0.3$; (a2)-(b2) $F/U=0.7$;
(a3)-(b3) $F/U=1$; The other system parameters are (a1)-(a3) $U=5\protect%
\kappa $, and (b1)-(b3) $U=10\protect\kappa $. It can be shown that the
resonant pulsed field can bring about the transition of the initial state
from the valence bond state to pairing state. The corresponding transfer
period is $\protect\pi /2\Delta $, which agrees with the theoretical result
in the main text. When the non-resonant external field is introduced, there
will still be some double-occupancy components in the evolved state, which
is beneficial to the conductivity of the system. In principle, the larger $U$%
, the larger the gap in the system and therefore more efficient this
transition. However, we can find that, by comparing the Figs. (a) and (b),
the efficiency of this dynamical scheme is still good even when a small $U$
is applied.}
\label{fig1}
\end{figure*}
\begin{figure}[tbp]
\centering
\includegraphics[width=0.45\textwidth]{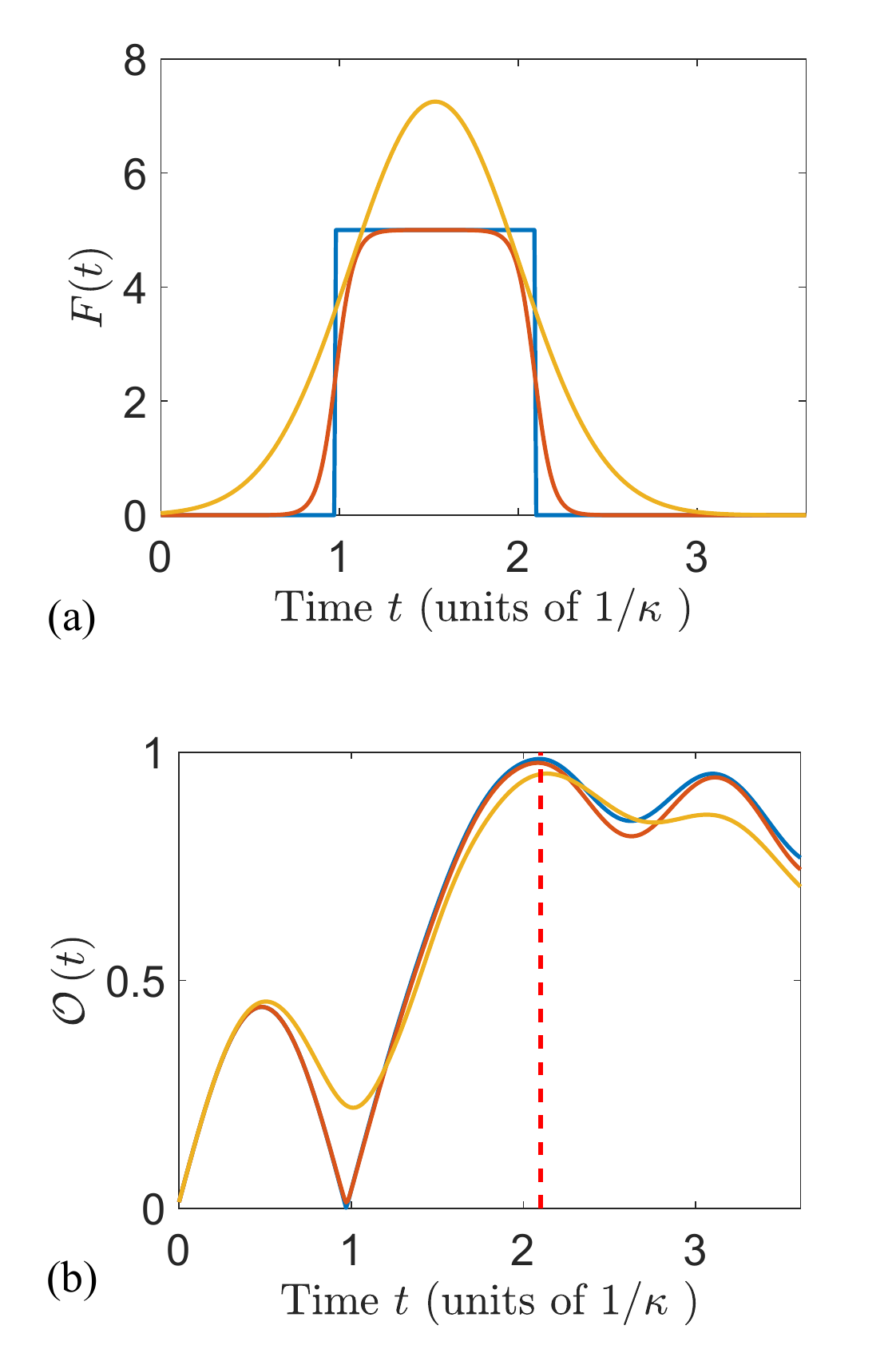}
\caption{Comparison of three typical
pulsed
fields.
The
system is
initialized in the valence bond state
with
$U=5\kappa $,
and
$F_{0}=U$. Fig.
\ref{fig1_plus} (a) plots the
shape of
$F\left(
t\right) $
and $F_{j}\left( t\right) $.
Here $F\left(
t\right)
$
represents a square
pulsed filed with $F\left(
t\right)
=F_{0}$
for
$T/2\leqslant t\leqslant
T$. The fidelity $\mathcal{O}\left(
t\right)
$
first oscillates because
$|\underline{2}\rangle $ is not
the
eigenstate
of
the system. When the
pulsed field is applied,
$\mathcal{O}\left(
t\right) $
approaches $1$. The
only difference between
such three
pulsed
fields is
the maximum value of
$\mathcal{O}\left(
t\right) $, which
is
indicated by
red dashed line. The
idea case
of
$\mathcal{O}\left(
t\right) =1$
requires: the interaction $U$
is
large
enough such that
$|\underline{2}\rangle $ is the
eigenstate
of
$H_{\mathrm{o}}$; the
resonant
pulsed filed $F_{0}=U$; the
exact
duration
time $T/2$. It can be
shown that
these two types of the
pulsed
filed can
fulfill the task that
excites the
system to the CDW
state
although
$F_{j}\left( t\right) $ does
not fully meet
these
conditions.}
\label{fig1_plus}
\end{figure}

\label{pairing} We start from the 1D Hubbard model subjected to a tilted
field, the Hamiltonian of which is given by
\begin{equation}
H=H_{\mathrm{o}}+H_{\text{\textrm{e}}},  \label{H}
\end{equation}%
with
\begin{eqnarray}
H_{\mathrm{o}} &=&-\kappa \sum_{j,\sigma }(c_{j,\sigma }^{\dagger
}c_{j+1,\sigma }+\text{H.c.})+U\sum_{j}n_{j,\uparrow }n_{j,\downarrow }, \\
H_{\text{\textrm{e}}} &=&F\sum_{j,\sigma }jn_{j,\sigma },
\end{eqnarray}%
where $c_{i,\sigma }$ ($c_{i,\sigma }^{\dagger }$) is the annihilation
(creation) operator for an electron at site $i$ with spin $\sigma \left(
=\uparrow ,\downarrow \right) $, and $n_{i,\sigma }=c_{i,\sigma }^{\dagger
}c_{i,\sigma }$. $\kappa $ is the hopping integral between the
nearest-neighboring sites, while $U>0$ is the on-site repulsive interaction.
To gain further insights into the field-induced paring, we first analyze the
symmetry of the system.
When the titled field is switched off, the system $H_{\mathrm{o}}$ respects the spin symmetry characterized by the
generators\begin{eqnarray}
s^{+} &=&\left( s^{-}\right) ^{\dagger }=\sum_{j}s_{j}^{+}, \\
s^{z} &=&\sum_{j}s_{j}^{z},
\end{eqnarray}where the local operators $s_{j}^{+}=c_{j,\uparrow }^{\dagger
}c_{j,\downarrow }$ and $s_{j}^{z}=\left( n_{j,\uparrow }-n_{j,\downarrow
}\right) /2$ obey the Lie algebra, i.e., $[s_{j}^{+},$ $s_{j}^{-}]=2s_{j}^{z} $, and $[s_{j}^{z},$ $s_{j}^{\pm }]=\pm s_{j}^{\pm }$.
Because of the commutation relation $[H_{\mathrm{o}}$, $\eta ^{+}]=U\eta
^{+} $, the system has a set of eigenstates generated by the $\eta $-pairing
operators, i.e., \{$\left( \eta ^{+}\right) ^{N_{\eta }}|\mathrm{Vac}\rangle
$\} where $|\mathrm{Vac}\rangle $ is a vacuum state with no electrons and $N_{\eta }$ is the number of $\eta $ pairs. Here, $\eta $ operator can be
explicitly written down as
\begin{eqnarray}
\eta ^{+} &=&\left( \eta ^{-}\right) ^{\dagger }=\sum_{j}\eta _{j}^{+}, \\
\eta ^{z} &=&\sum_{j}\eta _{j}^{z},
\end{eqnarray}with $\eta _{j}^{+}=\left( -1\right) ^{j}c_{j,\uparrow }^{\dagger
}c_{j,\downarrow }$ and $\eta _{j}^{z}=\left( n_{j,\uparrow
}+n_{j,\downarrow }-1\right) /2$ satisfying commutation relation, i.e., $[\eta _{j}^{+},$ $\eta _{j}^{-}]=2\eta _{j}^{z}$, and $[\eta _{j}^{z},$ $\eta _{j}^{\pm }]=\pm \eta _{j}^{\pm }$. At half-filling, the
ground state (GS) of $H_{\mathrm{o}}$ resides in the subspace with quantum
number $s^{2}=0$, $s^{z}=0$, and is often refereed to as the
anti-ferromagnetic ground state in the large $U$ limit ($U/\kappa \gg 1$).
It mainly consists of the Neel state. To give further insight into the pairing mechanism, we consider a two-site system, wherein the GS becomes a single valence bond state with
the form of $(c_{1,\uparrow }^{\dagger }c_{2,\downarrow }^{\dagger
}-c_{1,\downarrow }^{\dagger }c_{2,\uparrow }^{\dagger })/\sqrt{2}|\mathrm{%
Vac}\rangle $. The presence of $H_{\text{\textrm{e}}}$ does not break the
first spin symmetry but change the property of the GS. What we are
interested in is how does the system response to the tilted field if the
system is initialized in the GS of $H_{\mathrm{o}}$. For clarity, the matrix
form of Hamiltonian (\ref{H}) is written as
\begin{equation}
H=\left(
\begin{array}{ccc}
U+2F & -\sqrt{2}\kappa & 0 \\
-\sqrt{2}\kappa & 3F & -\sqrt{2}\kappa \\
0 & -\sqrt{2}\kappa & U+4F%
\end{array}%
\right) ,
\end{equation}%
in the invariant subspace $s^{2}=0$, $s^{z}=0$ under the basis of \{$|\underline{j}\rangle $\}, where
\begin{eqnarray}
|\underline{1}\rangle  &=&c_{1,\uparrow }^{\dagger }c_{1,\downarrow
}^{\dagger }|\mathrm{Vac}\rangle , \\
|\underline{2}\rangle  &=&\frac{1}{\sqrt{2}}(c_{1,\uparrow }^{\dagger
}c_{2,\downarrow }^{\dagger }-c_{1,\downarrow }^{\dagger }c_{2,\uparrow
}^{\dagger })|\mathrm{Vac}\rangle , \\
|\underline{3}\rangle  &=&c_{2,\uparrow }^{\dagger }c_{2,\downarrow
}^{\dagger }|\mathrm{Vac}\rangle .
\end{eqnarray}The presence of tilted field $F$ modulates the energy gap between the three
bases such that the system can exhibit rich dynamic behavior in addition to
doublon hopping in the large $U$ limit. Specifically, when we choose the
resonant field, that is, $F=U$, the energies of states $|\underline{1}%
\rangle $ and $|\underline{2}\rangle $ are close to resonance, but there is
an energy gap $2F$ between them and $|\underline{3}\rangle $.
Hence, one can
envisage that the evolved state will only oscillate periodically with
respect to two such bases if the system is initialized in the valence bond
state $|\underline{2}\rangle $. For simplicity, we sketch the effect of the resonant $F$ in Fig. \ref{fig_illustration_resonant}. Correspondingly, the propagator can be given
as $U=e^{i\sigma _{x}\Delta t}$ in the basis of \{$|\underline{1}\rangle $, $|\underline{2}\rangle $\}, and the transfer period is $T/2$, where $T=\pi
/\Delta $ with $\Delta =\sqrt{2}\kappa $. Fig. \ref{fig1} is plotted to
exhibit this transfer process with the initial state being valence bond
state, which agrees with the theoretical prediction. In the
experiment, the considered square pulsed field is not easy to realize due to its sharp
transition with time. For more realistic fields that vary slowly with time,
one can also arrive at the same result by carefully modulating parameters. As the examples,
we consider two different types of $F_{j}\left( t\right) $ ($j=1,$ $2$)
possessing the smoothed forms of
\begin{eqnarray}
F_{1}\left( t\right)  &=&\frac{F_{0}}{2}[\mathrm{thanh}\frac{\left(
t-T/2\right) }{\delta }-\mathrm{thanh}\frac{\left( t-T\right) }{\delta }], \\
F_{2}\left( t\right)  &=&1.45F_{0}e^{-\alpha ^{2}\left( t-3T/4\right) ^{2}},
\end{eqnarray}with $\delta =0.1$ and $\alpha =4\left( \text{\textrm{ln}}2\right) ^{1/2}/T$. $F_{0}$ is assumed to be equal to $U$. Here $\delta $ controls the slope of the curve on both sides and the
half-width of the Gaussian pulsed field $F_{2}\left( t\right) $ is assumed
to be $T/4$ such that it can excite the system to the CDW state. To check
the effect of these two realistic fields, the fidelity $\mathcal{O}\left(
t\right) =|\langle \underline{1}|e^{-iHt}|\underline{2}\rangle |$ is
introduced, where $|\underline{2}\rangle $ is the initial valence bond state
and $|\underline{1}\rangle $ is the target double-occupied state. Fig. \ref{fig1_plus}
shows clearly that $F_{j}\left( t\right) $ plays the same effect as that of
square pulsed field $F\left( t\right) $. So far we have demonstrated that the resonant titled field can transfer the
GS of $H_{\mathrm{o}}$ to a doublon state. The key point lies that $F$
places such two states on the same energy shell.

When we consider a Peierls distorted chain such that the nearest-neighbor hopping of $H_{\mathrm{o}}$ is staggered, the GS still has quantum number $s^{2}=0$, and $s_{z}=0$ \cite{Lieb1989}. However, the strong dimerization and large $U$  prescribe that GS is the direct product of a single valence bond in each dimerized unitcell forming a valence bond solid. This guarantees that the pulsed field takes effect in each unitcell so that the double-occupied states can be prepared individually with the same duration time $T/2$. As a consequence, the system is excited to the CDW state residing in the high energy sector. We sketch this process in Fig. \ref{fig_ill} for clarity.  This dynamical process plays a vital role in the formation of the non-equilibrium superconducting state. In the later section, we will show
that such a state can develop into a superconducting state.

\subsection{doublon dynamics}

\begin{figure*}[tbp]
\centering
\includegraphics[width=0.9\textwidth]{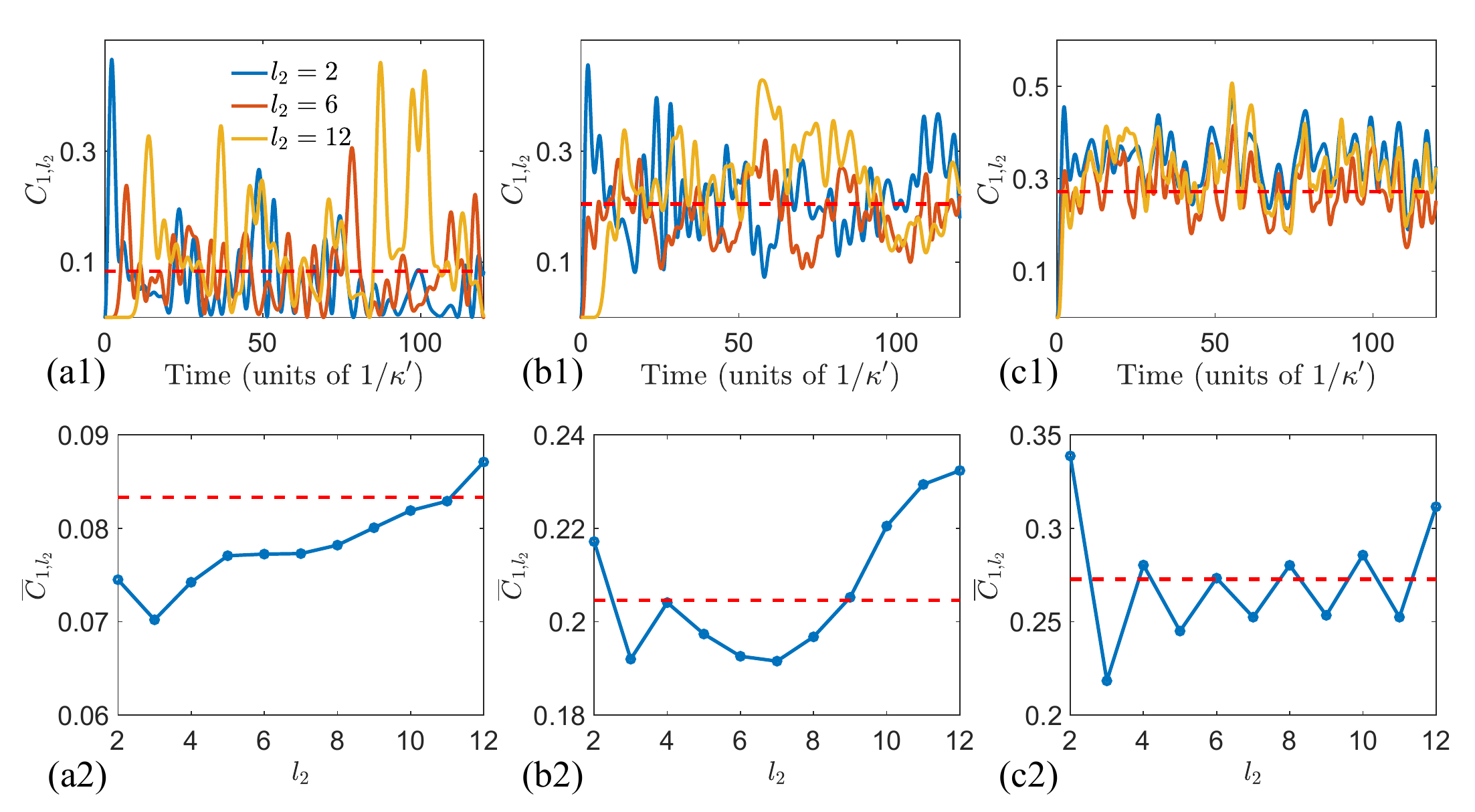}
\caption{(a1)-(c1) Time evolution of the correlations $C_{l_{1},l_{2}}\left(
t\right) $ for $2$, $6$ and $12$ filled particles of $12$-site Hubbard
model. (a2)-(c2) The averaged doublon-doublon correlators $\overline{C}%
_{1,l_{2}}$ as function of $l_{2}$. For simplicity, $l_{1}$ is assumed to be
$1$ and $l_{2}$ takes the values of $2$, $6$, and $12$ for (a1)-(c1),
respectively. The initial states are chosen as $|\protect\psi \left(
0\right) \rangle =\protect\eta _{1}^{+}|\mathrm{Vac}\rangle $, $\protect\eta %
_{1}^{+}\protect\eta _{3}^{+}\protect\eta _{5}^{+}|\mathrm{Vac}\rangle $,
and $\protect\eta _{1}^{+}\protect\eta _{3}^{+}\protect\eta _{5}^{+}\protect%
\eta _{7}^{+}\protect\eta _{9}^{+}\protect\eta _{11}^{+}|\mathrm{Vac}\rangle
$. The red lines of (a)-(c) obtained by Eq. (\protect\ref{Cr}) serve as the
benchmark to show whether the system is in the non-equilibrium
superconductivity phase. It is shown that the correlator first quickly
approaches the value of Eq. (\protect\ref{Cr}) and then it oscillates around
the red line. Such dynamic behavior is independent of $l_{2}$, which
indicates that the system reaches the superconducting phase featured by the
emergence of the steady state with ODLRO. The unit of time $t$ is the
inverse effective hopping rate $1/\protect\kappa ^{\prime }$ and the
duration time $\protect\tau $ of Figs. \protect\ref{fig2}(a2)-(c2) is
assumed as $240/\protect\kappa ^{\prime }$.}
\label{fig2}
\end{figure*}
\label{doublon} The dynamics of a spatially extended system of strongly
correlated fermions poses a notoriously complex many-body problem that is
hardly accessible to exact analytical or numerical methods.
In this section, we first study the single doublon dynamics in a uniform Hubbard model,
which may shed light on multi-doublon dynamics in the subsequently proposed
scheme. To begin with, we assume that the two fermions are initially at the
same site $j_{0}$, i.e., $|\psi _{i}\rangle =c_{j_{0},\uparrow }^{\dagger
}c_{j_{0},\downarrow }^{\dagger }|\mathrm{Vac}\rangle $. Two fermions
occupying the same site with strongly repulsive interaction $U$ form a
doublon manifested by the fact that the total double occupancy $%
D=\sum_{j}\langle D_{j}\rangle $ stays near $1$. The corresponding local
double-occupation operator is given by $D_{j}=n_{j,\uparrow }n_{j,\downarrow
}$. It is a long-living excitation, the decay of which is suppressed because
of energy conservation \cite{Hofmann2012}. Hence, in the large $U$ limit,
the doublon dynamics can be fully captured by the following effective $\eta $%
-spin model, in powers of $\kappa /U$:
\begin{equation}
H_{\mathrm{eff}}=-\kappa ^{\prime }\sum_{j}(\bm{\eta }_{j}\cdot \bm{\eta }%
_{j+1}-\frac{1}{4})  \label{heff}
\end{equation}%
which is obtained by a unitary transformation to project out the
energetically well separated high-energy part of the spectrum \cite%
{Fazekas1999}.
In its essence, a small cluster is enough to capture the feature of doublon movement and doublon-doublon interaction due to the absence of the long-range coupling. One can safely extend the result to a large system. In Appendix A, a simple two-site
case is provided to elucidate this mechanism. Note that we neglect the energy base $mU$ compared with Eq. (\ref{HH_eff}) in Appendix.
Here, $\kappa ^{\prime }=4\kappa ^{2}/U$ and $\bm{\eta }_{j}=(\eta _{j}^{x},$
$\eta _{j}^{y},$ $\eta _{j}^{z})$. For the repulsive interaction, $H_{%
\mathrm{eff}}$ describes a $\eta $-spin ferromagnetic model, the ground
state of which is $\eta $-pairing state with the form of $|\psi _{\mathrm{eff%
}}^{\mathrm{g}}\left( M\right) \rangle =\left( \eta ^{+}\right) ^{M}/\sqrt{%
\Omega }|\mathrm{Vac}\rangle $ where $M$ denotes the filled number of
doublons and the normalization efficient is $\Omega =C_{N}^{M}$.
The discussion about the uniform Hubbard model is
instructive for the effective Hamiltonian based on the dimerized Hubbard
model in the subsequent section since dimerization does not alter the
property of GS according to Lieb theorem \cite{Lieb1989}. It is worthy
pointing out that such paring ground state usually relates to the
superconductivity of the system due to the following $j$-independent
correlation relation \cite{Tindall2019,Yang1989}
\begin{equation}
|\langle \psi _{\mathrm{eff}}^{\mathrm{g}}\left( M\right) |\eta _{i}^{+}\eta
_{i+j}^{-}|\psi _{\mathrm{eff}}^{\mathrm{g}}\left( M\right) \rangle
|=\left\{
\begin{array}{c}
\frac{M(N-M)}{N(N-1)}\text{, for }j\neq 0 \\
\frac{M}{N}\text{, for }j=0%
\end{array}%
\right. .  \label{Cr}
\end{equation}%
It is also served as the building block to realize ODLRO state in the
subsequent non-equilibrium dynamic scheme. To gain further insight, we first
focus on the single-doublon case such that $\sum_{j}\eta _{j}^{z}\eta
_{j+1}^{z}$ only provides an energy base and plays no effect on the
dynamics. Hence, Eq. (\ref{heff}) takes the form of the tight-binding model
with the effective hopping $-\kappa ^{\prime }/2$, that is
\begin{equation}
H_{\mathrm{eff}}=-\frac{\kappa ^{\prime }}{2}\sum_{j}(\eta _{j}^{+}\eta
_{j+1}^{-}+\eta _{j}^{-}\eta _{j+1}^{+}).  \label{heff_single}
\end{equation}%
Performing the open boundary condition, the resulting free tight-binding
Hamiltonian is diagonalized by
a simple transformation (see Appendix B for more details). According to the Appendix B, one can readily
obtain the evolved state as
\begin{equation}
|\psi \left( t\right) \rangle =\sum_{j}g\left( j_{0},j,t\right) \eta
_{j}^{+}|\mathrm{Vac}\rangle ,
\end{equation}%
with
\begin{equation}
g\left( j_{0},j,t\right) =\sum_{l=-\infty }^{\infty }i^{C_{l.j_{0},j}^{1}}%
\mathcal{J}_{C_{l.j}^{1}}\left( 2\kappa ^{\prime }t\right)
-i^{C_{l.j_{0},j}^{2}}\mathcal{J}_{C_{l.j}^{2}}\left( 2\kappa ^{\prime
}t\right) ,
\end{equation}%
and%
\begin{eqnarray}
C_{l.j_{0},j}^{1} &=&j-j_{0}+lN+1, \\
C_{l.j_{0},j}^{2} &=&j-1+j_{0}+\left( l-1\right) N,
\end{eqnarray}%
where $\mathcal{J}_{l}$ denotes the $l$th Bessel function of the first kind.
We concentrate on the property of the evolved state after a long time scale.
To this end, two physical quantities are employed to characterize $|\psi
\left( t\right) \rangle $. The first is the expectation value of $%
d_{l}^{\dagger }d_{l}$ which can be given as
\begin{equation}
\overline{D}_{l}=\frac{1}{\tau }\int_{0}^{\tau }D_{l}\left( t\right) \mathrm{%
d}t\text{,}
\end{equation}%
where $D_{l}\left( t\right) =\left\langle \psi \left( t\right) \right\vert
d_{l}^{\dagger }d_{l}|\psi \left( t\right) \rangle $ represents the doublon
occupancy per site and $d_{l}^{\dagger }=c_{l,\uparrow }^{\dagger
}c_{l,\downarrow }^{\dagger }$. Here $\tau $ characterizes the relaxation
time that the system reaches to the steady state.
The second is the averaged
doublon-doublon correlation
\begin{equation}
\overline{C}_{l_{1},l_{2}}=\frac{1}{\tau }\int_{0}^{\tau
}C_{l_{1},l_{2}}\left( t\right) \mathrm{d}t\text{,}
\end{equation}where $C_{l_{1},l_{2}}\left( t\right) =|\left\langle \psi \left( t\right)
\right\vert \eta _{l_{1}}^{\dagger }\eta _{l_{2}}|\psi \left( t\right)
\rangle |$. Note that the Eq. (\ref{Cr}) can be employed as a benchmark to
examine whether the system reaches the superconductivity. Straightforward
algebra shows that $\overline{D}_{l}=1/\left( N+1\right) $ that is
irrelevant to the location of the initial state $j_{0}$. It also indicates
that the doublon is evenly distributed on each lattice site. Hence, one can
expect that $\overline{C}_{l_{1},l_{2}}$ is independent of the relative
distance between the two doublons. Due to the complexity of the analytical
solution of $\overline{C}_{l_{1},l_{2}}$, we fix $l_{1}=1$ and examine the
value of $\overline{C}_{1,l_{2}}$ as a function of $l_{2}$ in Fig. \ref{fig2}%
. It is shown that $\overline{C}_{1,l_{2}}$ does not depend on $l_{2}$. As
the increase of the number of the doublons, the correlator $\overline{C}%
_{1,l_{2}}$ still stays at a constant value, that is almost $\overline{D}%
_{l}=M/\left( N+1\right) $, manifesting that the result is not only
applicable to the case of dilute doublon gas. Figs. \ref{fig2}(a1)-(c1)
clear show that the correlations oscillate around the red lines. The values
of those lines are $0.0833$, $0.2045$, and $0.2727$, respectively, which are
obtained by setting $M=1$, $3$, $6$ and $N=12$ in the Eq. (\ref{Cr}).  This
indicates that such a non-equilibrium system can favor the existence of the
steady ODLRO state $|\psi _{\mathrm{eff}}^{\mathrm{g}}\left( M\right) \rangle $ on a long time scale in which $\eta $-pairing mechanism plays a vital role. Such a feature is exactly what makes the system
superconducting. In addition, we can also see that the system needs a
certain relaxation time to enter into the non-equilibrium superconducting
phase in Fig. \ref{fig2}. In such a dynamic process, the doublons gradually
diffuse throughout the whole lattice and finally forms a stead state with a
long-range correlation manifested by the oscillation of the correlator
around the red line. For the Hubbard model at half-filling in Fig. \ref{fig2}%
(c), one can roughly infer that such duration time is $4N/\kappa ^{\prime }$%
, which will be used to estimate the time scale of the subsequent dynamical
scheme. So far we have demonstrated the dynamic mechanism that can generate
the superconducting state from the Mott insulator phase via pulsed field. In
what follows, we will propose a scheme to prepare the ODLRO state based on
the dimerized Hubbard model.

\section{Scheme to preparing and detecting the ODLRO state}

\label{scheme} In this section, we concentrate on how to generate the ODLRO
state via out-of-equilibrium dynamics based on the SSH Hubbard model.
Further, we propose a dynamic method of detecting such non-equilibrium
superconducting phase.

\subsection{Dynamical preparing of the ODLRO state}

\label{generation}
\begin{figure*}[tbp]
\centering
\includegraphics[width=0.95\textwidth]{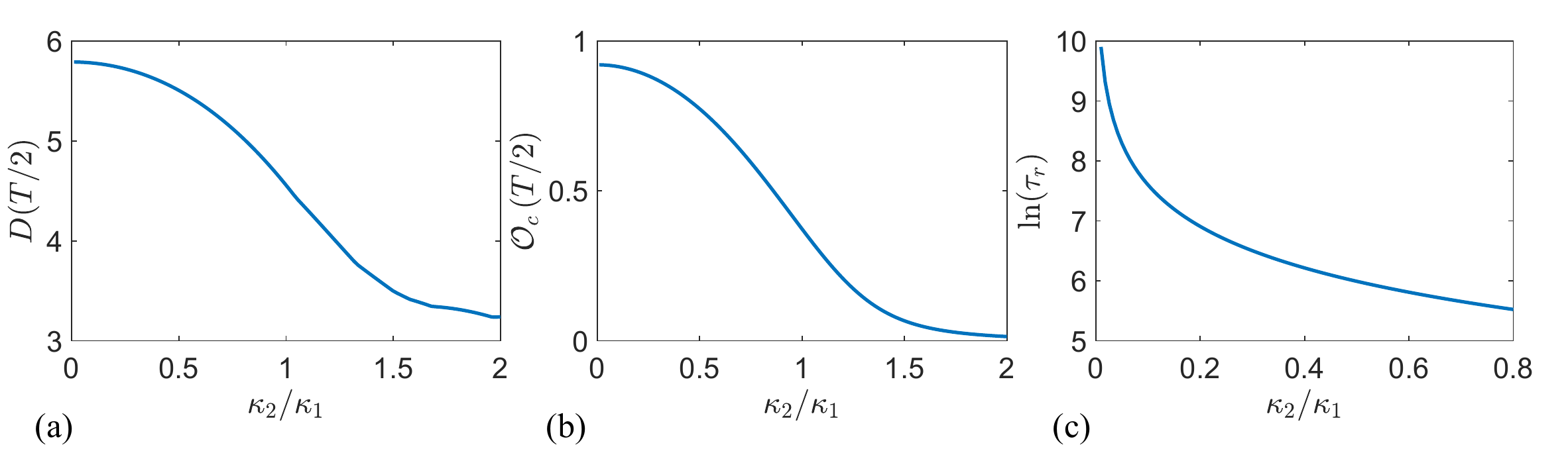}
\caption{Plots of $D\left(
T/2\right)
$,
$\mathcal{O}_{c}\left(
T/2\right) $, and $\tau _{r}$ as a
function
of
$\kappa _{2}/\kappa _{1}$ for
the system $H^{\mathrm{d}}$
with $12$
sites.
The system is initialized in
the GS of a half-filled
dimerized
Hubbard
model. The system parameters are $U=5\kappa _{1}$, and
$F=U$. It is
shown
that the increase of $\kappa
_{2}/\kappa _{1}$
decreases
$D\left(
T/2\right) $, and $\mathcal{O}_{c}\left( T/2\right)
$,
respectively. It
indicates that the initial state
cannot be excited to
the
high-energy
sector of $H_{\mathrm{o}}^{\mathrm{d}}$
such that
the
effective Hamiltonian
$H_{\mathrm{eff}}^{\mathrm{d}}$ does not
hold
for
the evolved state when
$t>T/2$. The degree of dimerization
affects
the
relaxation time $\tau
_{r}$. The stronger the dimerization
the longer
the
relaxation time. Hence,
one needs to choose a suitable
intercell
coupling
to generate a steady
superconducting state.}
\label{fig3_plus}
\end{figure*}

According to the two dynamic mechanisms proposed above, we will give the
method of generating ODLRO state through the dimerized Hubabrd model.
The considered $1$D time-dependent Hamiltonian can be given as
\begin{equation}
H^{\mathrm{d}}=H_{\mathrm{o}}^{\mathrm{d}}+H_{\mathrm{e}}^{\mathrm{d}},
\end{equation}where
\begin{eqnarray}
H_{\mathrm{o}}^{\mathrm{d}} &=&-\sum_{j=1}^{N-1}\sum_{\sigma =\uparrow
,\downarrow }(\kappa _{1}c_{2j-1,\sigma }^{\dagger }c_{2j,\sigma }  \notag \\
&&+\kappa _{2}c_{2j-1,\sigma }^{\dagger }c_{2j,\sigma }+\text{\textrm{H.c.}})
\notag \\
&&+U\sum_{j=1}^{2N}n_{j,\uparrow }n_{j,\downarrow },  \label{H_SSH} \\
H_{\mathrm{e}}^{\mathrm{d}} &=&F\left( t\right) \sum_{j=1}^{2N}\sum_{\sigma
=\uparrow ,\downarrow }jn_{j,\sigma },  \label{H_t}
\end{eqnarray}with
\begin{equation}
F\left( t\right) =\left\{
\begin{array}{c}
U,\text{ }0<t\leqslant T/2 \\
0,\text{ otherwise}\end{array}\right. .
\end{equation}When $U=0$, Eq. (\ref{H_SSH}) reduces to a celebrated Su-Schrieffer-Heeger model that is a
paradigm for characterizing the topology. Here, $\kappa _{1}/\kappa _{2}$
ratio controls the type of dimerization. In the OBC, we concentrate on the
GS property of $H_{\mathrm{o}}^{\mathrm{d}}$ and do not concern the edge
state behavior. Considering $H_{\mathrm{o}}^{\mathrm{d}}$ at half-filling,
the GS $|\psi _{\mathrm{d}}^{\mathrm{g}}\rangle $ of $H_{\mathrm{o}}^{%
\mathrm{d}}$ possesses the dimerized behavior if $\kappa _{1}>\kappa _{2}$,
which can be shown in Fig. \ref{fig_ill}. However, two end sites are not
paired if $\kappa _{1}<\kappa _{2}$. In the extreme case of $\kappa
_{1}/\kappa _{2}\gg 1$, the GS is fully dimerized and become a valence bond
solid. When the tilted field is applied, each dimerized sector respects the
dynamical mechanism developed in the subsection \ref{pairing}. As a
consequence, the valence bond solid state $|\Phi \left( 0\right) \rangle
=|\psi _{\mathrm{d}}^{\mathrm{g}}\rangle $ will evolve to a CDW state, that
is $|\Phi \left( T/2\right) \rangle =|\psi _{\mathrm{CDW}}\rangle $.
For clarity, this dynamical behavior is illustrated in Fig. \ref{fig_ill}. This is
the law that the evolved state should follow in an ideal case. In practice,
one can neither cut off the inter-cell coupling nor increase the on-site
interaction to the infinity. It can be envisioned that the presence of the
inter-cell coupling suppresses the dimerization and hence leads to the reduction of the component of $|\psi _{\mathrm{CDW}}\rangle $ in the evolved state. To check this point, we plot
the overlap $\mathcal{O}_{c}\left( t\right) =|\langle \Phi \left( t\right)
|e^{-iH^{\mathrm{d}}t}|\psi _{\mathrm{CDW}}\rangle |$ for different values
of $\kappa _{2}/\kappa _{1}$ in Fig. \ref{fig3_plus}. It is shown that $D\left(
T/2\right) $ and $\mathcal{O}_{c}\left( T/2\right) $ decrease as
intercell-coupling increases. This indicates that the GS is not excited to
the high-energy sector even though the resonant tilted field $F$ is applied.\textbf{\ }To ensure the success of the proposed scheme, one needs to choose
a small inter-cell coupling such that the quantity $\mathcal{O}_{c}\left( t\right) $ approaches $1$. However, a small enough $\kappa _{2}$
also brings other drawbacks, which will be seen later.

\begin{figure}[tbp]
\centering
\includegraphics[width=0.4\textwidth]{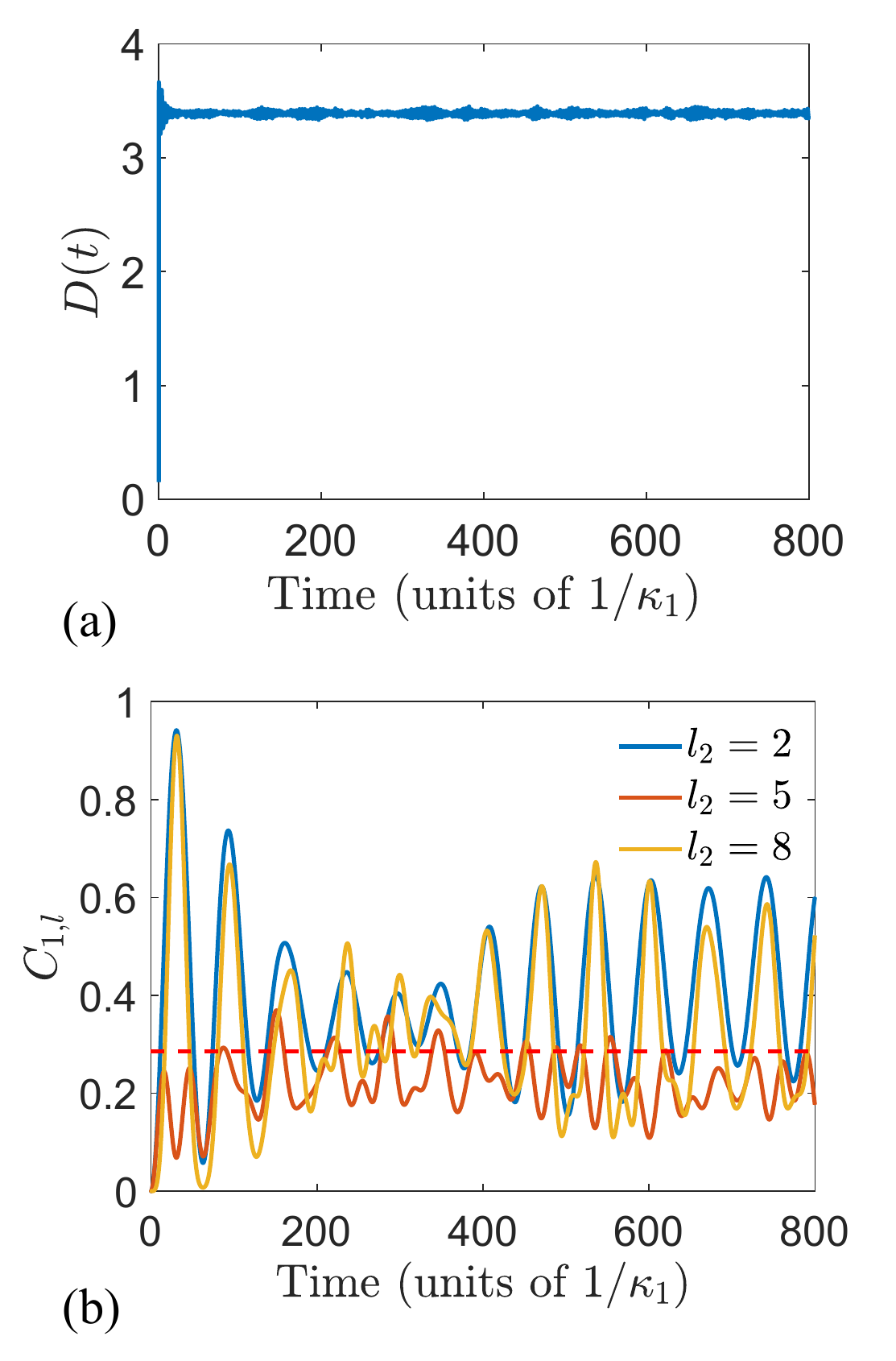}
\caption{(a1)-(c1) Upper panel: Time evolutions of total double occupancy $%
D\left( t\right) $. Lower panel: Time evolutions of $\protect\eta $
correlators $C_{1,2}\left( t\right) $, $C_{1,5}\left( t\right) $, and $%
C_{1,8}\left( t\right) $. The red line denotes the correlation of $\protect%
\eta $-pairing state $|\protect\psi _{\mathrm{eff}}^{\mathrm{g}}\left(
M\right) \rangle $ with $M=4$, which is served as a benchmark. The system is
initialized in the ground state of $H_{\mathrm{o}}^{\mathrm{SSH}}$ that is a
$8$-site $1$D SSH Hubbard model at half filling with $U=10\protect\kappa %
_{1} $, and $\protect\kappa _{2}=0.5\protect\kappa _{1}$. Under the action
of the resonant pulsed field, $D\left( t\right) $ first increases rapidly to
the peak, and then decays to $3.6$ followed by an almost constant trend. The
tiny fluctuations around the constant \textquotedblleft
final\textquotedblright\ value are simply reflecting the fact that the total
double occupancy does not commute with the Hamiltonian. Such the constant
value is determined by the degree of the dimerization and guarantees the
validity of our analytical analysis in the main text. After the quench, the
evolved state acquires the long-range correlation in the sense that the
correlator $C_{1,8}\left( t\right) $ oscillates around $0.3$. Note that the
time $t$ is measured in units of the inverse hopping $1/\protect\kappa _{1}$%
. }
\label{fig3}
\end{figure}

When $t>T/2$, the dynamics of $|\Phi \left( t>T/2\right) \rangle $ is only
govern by $H_{\mathrm{o}}^{\mathrm{d}}$. According to the mechanism shown in
the subsection \ref{doublon}, one can expect that the system will drive the
CDW state into a ODLRO state. The only difference lies in the effective
Hamiltonian regarding the doublon dynamics. It is a dimerized instead of a
uniform $\eta $-spin model that can be given as
\begin{eqnarray}
H_{\mathrm{eff}}^{\mathrm{d}} &=&-\kappa _{1}^{\prime }\sum_{j=1}^{N-1}(%
\bm{\eta }_{2j-1}\cdot \bm{\eta }_{2j}-\frac{1}{4})  \notag \\
&&-\kappa _{2}^{\prime }\sum_{j=1}^{N-1}(\bm{\eta }_{2j}\cdot \bm{\eta }%
_{2j+1}-\frac{1}{4}),
\end{eqnarray}%
where $\kappa _{j}^{\prime }=4\kappa _{j}^{2}/U$. However, such staggered
coupling coefficients do not alter the magnetic property of the system and
hence the corresponding ground state is still a $\eta $-spin ferromagnetic
state. This minor difference does not change the final steady state but only
affects the relaxation time due to the inhomogeneous effective hopping $%
\kappa _{j}^{\prime }$ which prohibits the diffusion of the doublon in the
entire lattice.
For clarity, we plot the relaxation time $\tau _{r}$ as a
function of $\kappa _{2}/\kappa _{1}$ in Fig. \ref{fig3_plus}. Here the relaxation time
refers to the duration of time that the system experiences when physical
observables $\overline{C}_{l_{1},l_{2}}$, and $\overline{D}_{l}$ do not vary
with time. This time is in proportion to $U/\kappa _{2}^{2}$, which can be
understood by the effective Hamiltonian $H_{\mathrm{eff}}^{\mathrm{d}}$.
Although the strong dimerization can ensure the main component of the final
state is $|\psi _{\mathrm{CDW}}\rangle $, the relaxation time is much longer
than the condition of weak dimerization due to the effective intercell
hopping $U/\kappa _{2}^{2}$. Evidently, the relaxation time is infinite when
the system is fully dimerized ($\kappa _{2}=0$). Given all of that, the
formation of the non-equilibrium superconducting state is a trade-off. On
the one hand, the strong dimerization ($\kappa _{1}/\kappa _{2}\gg 1$)
ensures that the GS of $H_{\mathrm{o}}^{\mathrm{d}}$ mainly consists of the
valence bond solid. Therefore, the combination of pulsed field and dimerized
Hubbard model can evolve the initial ground state to a CDW state which paves
the way to preparing the non-equilibrium ODLRO state. However, the cost is
to significantly suppress the effective hopping between the different
dimerized unit cells leading to a very long relaxation time. On the other
hand, if one decreases the degree of dimerization to $\kappa _{1}\approx
\kappa _{2}$, the main constituent of the GS is the Neel state although the
system is still in the Mott insulating phase. Such an initial GS cannot be
driven to the CDW state even though a resonant pulsed field is applied.
Therefore, the non-equilibrium superconducting phase fails to achieve. In
this point of view, the selection of hopping coefficient is a tradeoff
between the efficiency of the proposed scheme and the duration time.

\begin{figure}[tbp]
\centering
\includegraphics[width=0.4\textwidth]{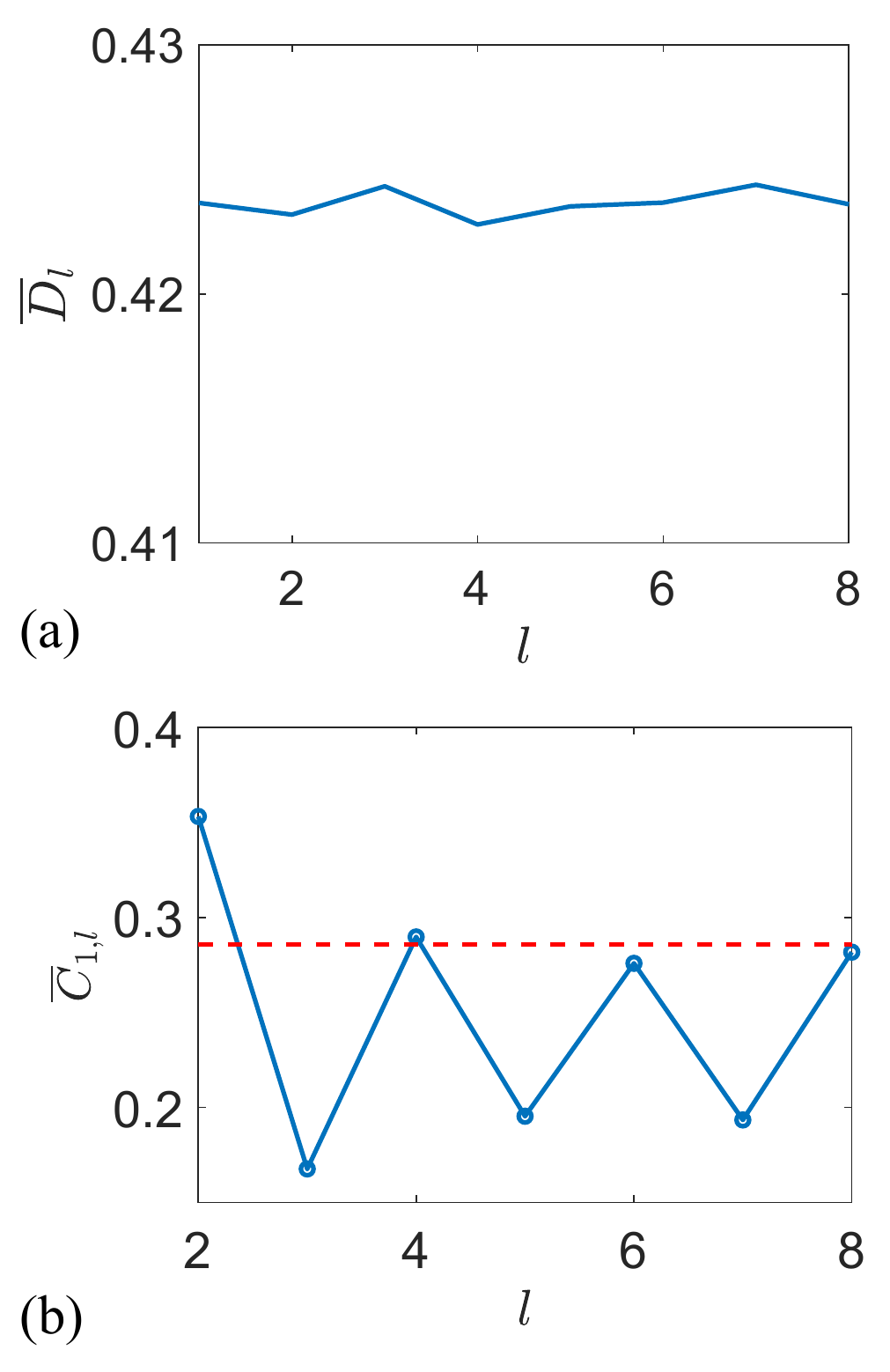}
\caption{Numerical results for the averaged $\overline{C}_{1,l}$ and $%
\overline{D}_{l}$. The corresponding duration time $\protect\tau $ is chosen
as $2000/\protect\kappa _{1}$. The other system parameters are the same as
those in Fig. \protect\ref{fig4}. $\overline{D}_{l}$ is evenly distributed
on each lattice site leading to $l$-independent $\protect\eta $-spin
correlation. Note that the red line denotes the correlation of the $\protect%
\eta $-pairing state $|\protect\psi _{\mathrm{eff}}^{\mathrm{g}}\left(
M\right) \rangle $ with $M=4$. Evidently, the steady superconducting state
is prepared via non-equilibrium dynamics.}
\label{fig4}
\end{figure}

In Fig. \ref{fig3}, we demonstrate this dynamical scheme by setting $\kappa
_{1}/\kappa _{2}=2$. In this setting, the portion of the valence-bond-solid
state in the GS is about $0.9$. Hence, after a resonant pulsed field, the
expectation value $\sum_{l}D_{l}\left( t\right) $ of the target state is
approximately $3.6$. Fig. \ref{fig3}(a) clearly shows that the total double
occupancy quickly approaches $3.6$ and is stabilized around that value
protected by the energy conservation. Such long-lived excitation guarantees
the validity of the effective $\eta $-spin ferromagnetic model in the
subsequent doublon-diffusion dynamics. Consequently, the long-range
correlation of $\eta $ spin is established as shown in Fig. \ref{fig3}(b). To
give a panoramic view of the dynamical scheme, we also perform the numerical
simulation in Fig. \ref{fig4} to show the time evolutions of $\overline{D}%
_{l}$ and $\overline{C}_{1,l_{2}}$. It can be shown that the final steady
state distributes evenly in the entire lattice with $\overline{D}%
_{l}=M/(N+1) $. This indicates the uniform diffusion of the doublons over
the lattice. Furthermore, the averaged correlator $\overline{C}_{1,l}$
oscillates around $0.28$ suggesting that the system enters into the
non-equilibrium superconducting phase, which verifies the previous analysis.
In experiment, the proposed scheme could be implemented in the ultracold
atoms loaded in optical lattices \cite{Juenemann2017,Abanin2019}. The tunability
and long coherence times of this system, along with the ability to prepare
highly nonequilibrium states, enable one to probe such quantum dynamics.

\subsection{Dynamical detection of the non-equilibrium superconducting phase}

\label{detection} 
\begin{figure}[tbp]
\centering
\includegraphics[width=0.4\textwidth]{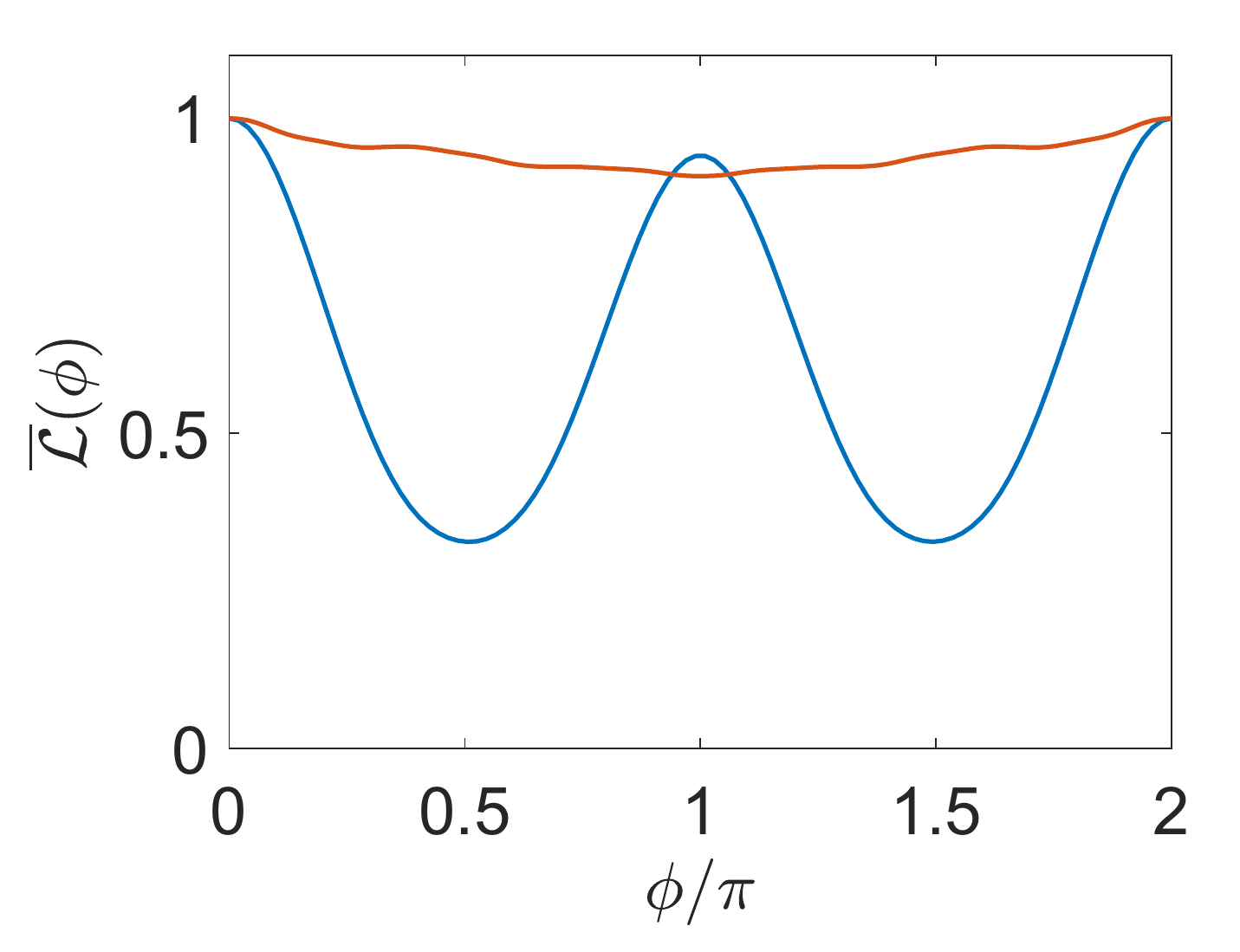}
\caption{The average LE
($\overline{\mathcal{L}}$)
for
different intercell
couplings with
resonant pulsed field $F$.
The
simulation is performed in a $12$-site
Hubbard model at half-filling
with
$s^{z}=0$, and the total
magnetic flux
penetrating the 1D ring is
taken as
$12\phi $. The red and
blue lines
denote $\kappa _{2}/\kappa
_{1}=0.3$, and
$1$, respectively. The
other
system parameters are
$U=10\kappa _{1}$, and
$t_{f}=1200/\kappa
_{1}$.
For the uniform chain,
the final evolved state
lies in the
low-energy
sector
of
$H_{\mathrm{p}}^{\mathrm{b}}$
and
$H_{\mathrm{o}}^{%
\mathrm{d}}$.
Correspondingly, the effective
Heisenberg
Hamiltonians of
both systems share
the same form as shown in
Eq.
(\ref{heff_s}). Hence,
the average LE ($\overline{\mathcal{L}}$) does
not
respond to the magnetic
flux. For the
strong dimerization, the
resonant
pulse field can fully
drive the
valence-bond-solid state into a CDW
state.
Its dynamics can be
captured by $\eta $-spin Hamiltonian
(\ref{heff_eta})
rather than
Heisenberg Hamiltonian (\ref{heff_s}). When a
doublon moves, it
acquires a
$2\phi $ phase factor
that can be witnessed
by the decrease
of
$\mathcal{L}\left( t\right) $.
However, when $\phi =\pi
$, the
two
effective Hammiltonians
$H_{\mathrm{eff}}^{\mathrm{p}}$
and
$H_{\mathrm{eff}}^{\mathrm{d}}$ are
the same so that
$\mathcal{L}\left(
t\right) $ returns to $1$. In this
sense, the
periodical
oscillation
behavior of
$\overline{\mathcal{L}}\left( \phi
\right) $ may
serve as a
dynamical
signature to probe whether the systems
enters into
the
non-equilibrium
superconducting phase.}
\label{fig_LE}
\end{figure}

To further capture the superconductivity of the
non-equilibrium system, we introduce the LE, which is a measure of
reversibility and sensitivity to the perturbation of quantum evolution. The
perturbation considered in our scheme is the magnetic flux threading the
ring. To this end, an additional quench process should be implemented.
The corresponding post-quench Hamiltonian can be given as
\begin{eqnarray}
H_{\mathrm{p}}^{\mathrm{b}} &=&-\sum_{j=1}^{N}\sum_{\sigma =\uparrow
,\downarrow }(\kappa _{1}e^{i\phi }c_{2j-1,\sigma }^{\dagger }c_{2j,\sigma }
\notag \\
&&+\kappa _{2}e^{i\phi }c_{2j,\sigma }^{\dagger }c_{2j+1,\sigma }+\text{\textrm{H.c.}})  \notag \\
&&+U\sum_{j=1}^{N}n_{j,\uparrow }n_{j,\downarrow },
\end{eqnarray}where $c_{2N+j,\sigma }=c_{j,\sigma }$, and $\phi =2N\phi $ denotes the
total magnetic flux piercing the ring. Taking the steady state $|\Phi \left(
t_{f}\right) \rangle $ as an initial state, the LE is defined as
\begin{equation}
\mathcal{L}\left( t\right) =|\langle \Phi \left( t_{f}\right) |e^{-iH_{\mathrm{p}}^{\mathrm{b}}t}e^{iH_{\mathrm{o}}^{\mathrm{b}}t}|\Phi \left(
t_{f}\right) \rangle |^{2},  \label{Lt}
\end{equation}where $t_{f}$ is relaxation time of the first quench dynamics. {Eq. (\ref{Lt}) represents the overlap at time $t$ of two states evolved from $\left\vert
\Phi \left( t_{f}\right) \right\rangle $ under the action of the Hamiltonian
operators }$H_{\mathrm{o}}^{\mathrm{b}}${\ and }$H_{\mathrm{p}}^{\mathrm{b}}${. }Consider a typical case $\kappa _{1}\sim \kappa _{2}$, the GS of $H_{\mathrm{o}}^{\mathrm{d}}$ at half-filling is an anti-ferromagnetic state.
The resonant pulsed field $H_{\mathrm{e}}^{\mathrm{d}}$ does not induce the
particle pairing and hence cannot place the evolved state $|\Phi \left(
t_{f}\right) \rangle $ in the high-energy sector of $H_{\mathrm{o}}^{\mathrm{d}}$. It is still an insulating state residing in the low-energy sector and its dynamics is
described by the effective Heisenberg Hamiltonian
\begin{eqnarray}
H_{\mathrm{eff}}^{\mathrm{s}} &=&-\kappa _{1}^{\prime }\sum_{j=1}^{N}(\bm{s}_{2j-1}\cdot \bm{s }_{2j}-\frac{1}{4})  \notag \\
&&-\kappa _{2}^{\prime }\sum_{j=1}^{N}(\bm{s}_{2j}\cdot \bm{s }_{2j+1}-\frac{1}{4}).  \label{heff_s}
\end{eqnarray}Because of the virtual exchange of particles, this Hamiltonian does hold
regardless of the presence or absence of the magnetic field. As a
consequence, the post- and before-quench Hamiltonians share the same
effective Hamiltonian $H_{\mathrm{eff}}^{\mathrm{s}}$ such that $\mathcal{L}\left( t\right) $ stay at $1$. Now we switch gear to another typical case $\kappa _{1}/\kappa _{2}\gg 1$ in which the steady state $|\Phi \left(
t_{f}\right) \rangle $ resides in the high-energy sector due to the resonant
pulsed field. It is a superconducting state featured by the constant $\eta $-spin correlator. With the same spirit, one can obtain the effective
post-quench Hamiltonian in such a sector when the magnetic field is applied.
According to the Appendix A, it can be given as\begin{eqnarray}
H_{\mathrm{eff}}^{\mathrm{p}} &=&-\frac{\kappa _{1}^{\prime }}{2}\sum_{j=1}^{N}(e^{i2\phi }\eta _{2j-1}^{+}\eta _{2j}^{-}+\text{\textrm{H.c.}}
\notag \\
&&+2\eta _{2j-1}^{z}\eta _{2j}^{z}-\frac{1}{2})  \notag \\
&&-\frac{\kappa _{2}^{\prime }}{2}\sum_{j=1}^{N}(e^{i2\phi }\eta
_{2j}^{+}\eta _{2j+1}^{-}+\text{\textrm{H.c.}}  \notag \\
&&+2\eta _{2j}^{z}\eta _{2j+1}^{z}-\frac{1}{2}),  \label{heff_eta}
\end{eqnarray}where the phase factor $e^{i2\phi }$ stems from the doublon hopping. This
ensures that the system can respond to the external magnetic field, and
hence $\mathcal{L}\left( t\right) $ changes. Note that when $\phi =n\pi $,
the effective post- and before-quench Hamiltonians are the same as each other
resulting in $\mathcal{L}\left( t\right) =1$. If we fix the reversal time $t=\tau $, the value of $\mathcal{L}\left( \tau \right) $ will show a
periodical behavior as $\phi $ varies. In this sense, whether the LE exhibits periodic behavior
is an important feature to mark whether the particles move in pairs. To confirm this conclusion, a
numerical simulation of average $\overline{\mathcal{L}}$ defined as\begin{equation}
\overline{\mathcal{L}}=\frac{1}{\tau }\int_{0}^{\tau }|\langle \Phi \left(
t_{f}\right) |e^{-iH_{\mathrm{p}}^{\mathrm{b}}t}e^{iH_{\mathrm{o}}^{\mathrm{b}}t}|\Phi \left( t_{f}\right) \rangle |^{2}\mathrm{d}t
\end{equation}is performed in Fig. \ref{fig_LE}. It is shown that when $\kappa _{2}/\kappa _{1}=0.3$, $\overline{\mathcal{L}}\left( \phi \right) $ exhibits an oscillation with
period $\phi =\pi $, which agrees with our prediction. On the contrary, $\overline{\mathcal{L}}\left( \phi \right) $ stays at $1$ if $\kappa
_{2}/\kappa _{1}=1$ indicating that system is still in the Mott insulating
phase. This scheme suggests an alternative dynamical approach to detecting
the non-equilibrium phase of matter.

\section{Summary}

\label{summary} In summary, we have proposed a non-equilibrium method to
realize the long-living superconductivity in the dimerized
Hubbard model. The underlying mechanism can be dissected into two main
dynamical processes, dynamical pairing and doublon dynamics in the
highly-excited subspace. Specifically, the dimerization in the Hubbard model
makes the main component of the anti-ferromagnetic ground state change from
Neel state to a valence bond solid. Therefore, the dynamical pairing is
confined to each unitcell such that the corresponding valence bond state is
excited to the doublon state forming the so-called CDW state. When the
external field is switched off, the energy conservation prevents the decay
of the doublon and hence protects such the long-lived excitation. The
dynamics of the CDW state is determined by the highly-excited state of the %
dimerized Hubbard model, which can be described by a
Heisenberg-like $\eta $-spin ferromagnetic model. After a long-time
evolution, the doublons tend to distribute evenly in the entire lattice and
form a steady state with ODLRO. Furthermore, we propose a dynamical
detection method to identify this non-equilibrium superconducting phase via
introducing the magnetic flux to trigger a quench and measuring the LE. Our
results open a new avenue toward enhancing and detecting superconductivity
through non-equilibrium dynamics.

\acknowledgments We acknowledge the support of the National Natural Science
Foundation of China (Grants No. 11975166, and No. 11874225).

\appendix\label{appendix}

\section{Simple example of two-site case for the effective Hamiltonian $H_{%
\mathrm{eff}}$}

In this section, our goal is to obtain the effective Hamiltonian (\ref{heff}). To this end, we first divide the Hamiltonian $H_{\mathrm{o}}$ into two
parts $H_{\mathrm{o}}=H_{0}+H_{t}$, where
\begin{eqnarray}
H_{0} &=&U\sum_{j=1}n_{j,\uparrow }n_{j,\downarrow }, \\
H_{t} &=&-\kappa \sum_{\sigma ,j}(c_{j,\sigma }^{\dagger }c_{j+1,\sigma }+\text{H.c.}).
\end{eqnarray}To second order in perturbation theory, the effective Hamiltonian is given by\begin{equation}
H_{\mathrm{eff}}=P_{0}H_{0}P_{0}+P_{0}H_{t}P_{1}\frac{1}{E_{0}-H_{0}}P_{1}H_{t}P_{0}+O\left( \frac{\kappa ^{3}}{U^{2}}\right) ,  \label{sheff}
\end{equation}where $P_{0}$ is a projector onto the Hilbert subspace in which there are $m$
lattice sites occupied by two particles with opposite spin orientation, and $P_{1}=1-P_{0}$ is the complementary projection. Here the energy $E_{0}$ of
the unperturbed state is set to $E_{0}=mU$ where $m$ denotes the number of
doublons. Since $H_{t}$ acting on states in $P_{0}$ annihilates only one
double occupied site, all states in $P_{1}H_{t}P_{0}$ have exactly $m-1$
doubly occupied sites. Now we provide a detailed calculation of the two-site
case for the effective Hamiltonian $H_{\mathrm{eff}}$ which may shed light
to obtain the effective Hamiltonian (\ref{heff}). In the simplest two-site
case, $P_{0}=\sum_{\alpha \in \text{\textrm{d.o.}}}|\alpha \rangle \langle
\alpha |$ is the projection operator to the doublon subspace spanned by the
configuration $\left\{ |\text{\textrm{x}}0\rangle ,|0\text{\textrm{x}}\rangle \right\} $, and $P_{1}=1-P_{0}=\sum_{a\notin \text{\textrm{d.o.}}}|a\rangle \langle a|$ is the complementary projection. Here the
abbreviation \textrm{d.o.} means the doubly occupied subspace and $|$\textrm{x}$0\rangle =c_{1,\uparrow }^{\dagger }c_{1,\downarrow }^{\dagger }|$\textrm{Vac}$\rangle $, $|0$\textrm{x}$\rangle =c_{2,\uparrow }^{\dagger
}c_{2,\downarrow }^{\dagger }|$\textrm{Vac}$\rangle $. The first term of Eq.
(\ref{sheff}) clear gives $P_{0}H_{0}P_{0}=U$. The second term can be
simplified by noting: (i) the unperturbed energy $E_{0}$ is $U$; (ii) $P_{1}H_{t}P_{0}$ annihilates the doubly occupied site. Then $H_{\mathrm{eff}}^{2}$ for two-site Hubbard system can be written as
\begin{eqnarray}
H_{\mathrm{eff}}^{2} &=&U+\sum_{\alpha ,\beta \in \text{d.o.}}\sum_{a,b\notin \text{d.o.}}|\alpha \rangle \langle \alpha |H^{\prime
}|a\rangle \langle a|  \notag \\
&&\times \frac{1}{U-H_{0}}|b\rangle \langle b|H^{\prime }|\beta \rangle
\langle \beta |  \notag \\
&=&U+\frac{1}{U}\sum_{\alpha ,\beta \in \text{d.o.}}\langle \alpha |\left(
H^{\prime }\right) ^{2}|\beta \rangle |\alpha \rangle \langle \beta |.
\end{eqnarray}The second term describes the virtual exchange of the fermions yielding that
\begin{equation}
H_{\mathrm{eff}}^{2}=U+\frac{2\kappa ^{2}}{U}\left( |\text{\textrm{x}}0\rangle \langle 0\text{\textrm{x}}|+|0\text{\textrm{x}}\rangle \langle
\text{\textrm{x}}0|+|\text{\textrm{x}}0\rangle \langle \text{\textrm{x}}0|+|0\text{\textrm{x}}\rangle \langle 0\text{\textrm{x}}|\right) .
\end{equation}Combining the cases in the subspaces of $|\mathrm{xx}\rangle $ and $|$\textrm{Vac}$\rangle $, the pseudo spin Hamiltonian can be given by the Heisenberg-like model
\begin{equation}
H_{\mathrm{eff}}=U-\frac{4t^{2}}{U}\left( \bm{\eta }_{1}\cdot \bm{\eta }_{2}-\frac{1}{4}\right) ,
\end{equation}where $\bm{\eta }_{j}=(\eta _{j}^{x},$ $\eta _{j}^{y},$ $\eta _{j}^{z})$,
and $m$ can be $0$, $1$, and $2$ denoting the number of pairs of the doublon
subspace. Evidently, the GS of $H_{\mathrm{eff}}$ is the $\eta $-spin
ferromagnetic state with the form of $\left( \eta ^{+}\right) ^{2}|$\textrm{Vac}$\rangle $. One can extend the result to the system with $N$ sites, the
corresponding effective Hamiltonian is given as
\begin{equation}
H_{\mathrm{eff}}=mU-\frac{4t^{2}}{U}\sum_{j}\left( \bm{\eta }_{j}\cdot \bm{\eta }_{j+1}-\frac{1}{4}\right) .  \label{HH_eff}
\end{equation}Hence, the ferromagnetic state of $\eta $ spins aligned on the $x-y$ plane
is the $\eta $-pairing superconducting state.

\section{The dynamics of a single doublon in a finite chain}

The diffusion of the doublon on the entire lattice is the key to achieving
the non-equilibrium superconducting phase of the proposed scheme. Here, we
give a single doublon dynamics analytically, which may shed light on dilute
doublon gas. Starting from effective Hamiltonian (\ref{heff_single}), it is
a free tight-binding Hamiltonian with open boundary condition, which can be
diagonalized by the following transformation\begin{eqnarray}
\eta _{k}^{+} &=&\sqrt{\frac{2}{N+1}}\sum_{j}\sin \left( kj\right) \eta
_{j}^{+}|\mathrm{Vac}\rangle , \\
\eta _{k}^{-} &=&\sqrt{\frac{2}{N+1}}\sum_{j}\sin \left( kj\right) \eta
_{j}^{-}|\mathrm{Vac}\rangle ,
\end{eqnarray}where $k=n\pi /(N+1)$. Correspondingly, the effective Hamiltonian in this
representation can be given as
\begin{equation}
H_{\mathrm{eff}}=\sum_{k}\varepsilon _{k}\eta _{k}^{+}\eta _{k}^{-}
\end{equation}with eigen energy $\varepsilon _{k}=-\kappa ^{\prime }\cos k$. Consider a
double-occupied initial state with form\begin{equation}
|\psi \left( t\right) \rangle =\eta _{j_{0}}^{+}|\mathrm{Vac}\rangle ,
\end{equation}, one can readily obtain the evolved state in terms of operator $\eta
_{k}^{+}$ as\begin{equation}
|\psi \left( t\right) \rangle =\sqrt{\frac{2}{N+1}}\sum_{k}e^{-i\varepsilon
_{k}t}\sin kj_{0}\eta _{k}^{+}|\mathrm{Vac}\rangle ,
\end{equation}Taking the inverse transformation, the evolved state in the coordinate space
is
\begin{equation}
|\psi \left( t\right) \rangle =\sum_{j}g\left( j_{0},j,t\right) \eta
_{j}^{+}|\mathrm{Vac}\rangle ,
\end{equation}where
\begin{equation}
g\left( j_{0},j,t\right) =\frac{2}{N+1}\sum_{k}e^{-i\varepsilon _{k}t}\sin
kj\sin kj_{0},  \label{p}
\end{equation}can be deemed as the propagator describing how much the probability of the
doublon flow from the initial $j_{0}$th to $j$th site. In the limit $N\rightarrow \infty $, the summation $\sum_{k}/N$ in Eq. (\ref{p}) can be
replaced by the integral $\int \mathrm{d}k$ such that
\begin{equation}
g\left( j_{0},j,t\right) =i^{j-j_{0}}\mathcal{J}_{j-j_{0}}\left( 2\kappa
^{\prime }t\right) -i^{j+j_{0}}\mathcal{J}_{j+j_{0}}\left( 2\kappa ^{\prime
}t\right) ,
\end{equation}where $\mathcal{J}_{l}$ denotes the $l$th Bessel function of the first
kind. However, such substitution is not true as $N$ is a finite number. As
an alternative, the summation in Eq. (\ref{p}) can be expanded by the Bessel
function as\begin{equation}
g\left( j_{0},j,t\right) =\sum_{l=-\infty }^{\infty }i^{C_{l.j_{0},j}^{1}}\mathcal{J}_{C_{l.j_{0},j}^{1}}\left( 2\kappa ^{\prime }t\right)
-i^{C_{l.j_{0},j}^{2}}\mathcal{J}_{C_{l.j_{0},j}^{2}}\left( 2\kappa ^{\prime
}t\right) ,  \label{g_in}
\end{equation}with
\begin{eqnarray}
C_{l.j_{0},j}^{1} &=&j-j_{0}+lN+1, \\
C_{l.j_{0},j}^{2} &=&j-1+j_{0}+\left( l-1\right) N.
\end{eqnarray}From another point of view, the dynamics in a finite chain can be obtained
by projecting the dynamics of an infinite system to such a finite system. In
this scenario, one can utilize safely the Bessel function to capture the
interference behavior when the evolved state touches the boundary. The
cost is to project the Bessel function entirely into the subsystem. The
infinite summation of Eq. (\ref{g_in}) denotes such a physical process.


\end{document}